\documentclass{article}
\usepackage{LaThuileFPSpro}
\newcommand{\ptjet}{\rm P_{\rm T}^{\rm jet}}

\newcommand{\D}       {\rm D0\hspace{-1.15ex}/\hspace{1.2ex}}
\begin{document}
\title{ 
  STUDIES OF JET PROPERTIES AT THE TEVATRON
  }
\author{
  Mario Mart\'\i nez        \\
  {\em Instituto de F\'\i sica de Altas Energ\'\i as} \\
   \em    Universidad Aut\'onoma de Barcelona \\
   \em    E-80193 Bellaterra (Barcelona) Spain \\
   (\em On behalf of the CDF Collaboration)
  }
\maketitle

\baselineskip=11.6pt

\begin{abstract}
In this contribution, a number of new QCD results on jet production 
from the CDF and $\D$ experiments in Run II are discussed in detail.  
\end{abstract}
\newpage
\section{Introduction}

The Run II at the Tevatron will define a new
level of precision for QCD studies in hadron collisions.
Both collider experiments, CDF and D0, expect
to collect up to $8 \  {\rm fb}^{-1}$ of data in this new
run period. The increase in instantaneous luminosity,
center-of-mass energy (from 1.8 TeV to 2 TeV) and the improved
acceptance of the detectors will allow stringent tests of the
Standard Model (SM) predictions in extended regions of jet transverse
momentum, $P^{\rm jet}_T$, and jet rapidity, $Y^{\rm jet}$. 
The hadronic final states in hadron-hadron collisions are characterized by the presence of
soft contributions (the so-called {\it{underlying event}}) from initial-state gluon radiation and 
mutiple parton interactions between remnants, in addition to the jets of hadrons originated by the hard interaction.
A proper comparison with pQCD predictions at the parton level requires an adequate modeling of these
soft contributions which become important at low $\ptjet$. 
In this letter, a review of some of the most important QCD results from
Run II  is presented.


\section{Inclusive Jet Production at the Tevatron}

The measurement of the inclusive jet production cross section for central 
jets constitutes one of the pillars of the jet physics program at 
the Tevatron. It provides a stringent test of perturbative QCD predictions 
over almost nine orders of magnitude and 
probes distances up to $\sim 10^{-19} {\rm m}$.  
Thanks to the increase in the center-of-mass energy in Run II 
the jet production rate has been multiplied (by a factor of five for jets with $P_T^{\rm jet} > 600$~GeV) 
and the first measurements have already extended the $P_T^{\rm jet}$ coverage 
by 150~GeV compared to Run I. In addition, both CDF and D0~experiments explore new 
jet algorithms following the theoretical work that indicates that the 
cone-based jet algorithm employed in Run I  
is not infrared safe and compromises a future 
meanful comparison with pQCD calculations at NNLO. 
\begin{figure}[h]
  \centerline{\hbox{ \hspace{-6 cm} 
    \special{epsfile=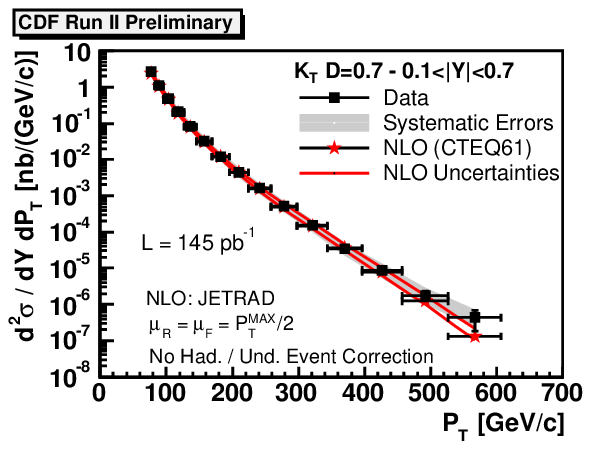 hscale=0.95 vscale=1.  }
\hspace{5.5 cm}
    \special{epsfile=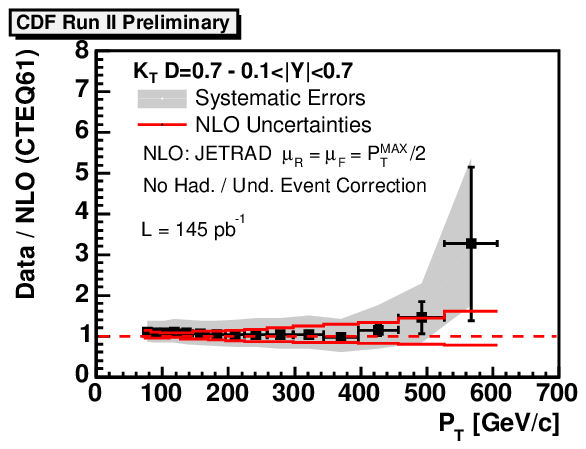 hscale=0.95 vscale=1.  }   
   }}
\vspace{4 cm}
 \caption{\it
       The measured inclusive jet cross section compared to pQCD NLO predictions. Jets 
are seached for using the longitudinally invariant $K_T$ algorithm. 
    \label{fig1} }
\end{figure}
Figure~\ref{fig1} shows the measured inclusive jet cross section by CDF using the 
longitudinally invariant $K_T$ algorithm \cite{soper} and based on the first 145 $\rm{pb}^{-1}$ of Run II data. 
Measurements have been performed using values for the D parameter in the $K_T$ expression,
\begin{equation}
K_{ij} = \rm{min}(p_{T,i}^2,p_{T,j}^2) \cdot \frac{(y_i - y_j)^2+(\phi_i - \phi_j)^2}{D} \ \ ,
\end{equation}
equal to 0.5, 0.7 and 1.0. 
The measurements are compared to pQCD NLO calculations~\cite{nlo} using  
CTEQ6~\cite{cteq} parton density functions in the proton and antiproton and the renormalization and  
factorization scales set to $p_T^{\rm max}/2$. The measured  cross section is reasonably  
well described by the predictions for $P_T^{\rm jet} > 150 {\rm \ GeV}$ within 
the present uncertainties. The systematic errors on the data are dominated by the 
uncertainty on the jet energy scale determination while the theoretical predictions 
suffer from our limited knowledge of the gluon distribution at high $x$. 
At lower $P_T^{\rm jet}$, the data is  
systematically above the predictions  and the effect increases  
as D increases (see Figure~\ref{fig2}). This indicates the presence of   
soft-gluon contributions and fragmentation effects    
that have not been taken into account yet. 
\begin{figure}[htbp]
  \centerline{\hbox{ \hspace{-6 cm}
    \special{epsfile=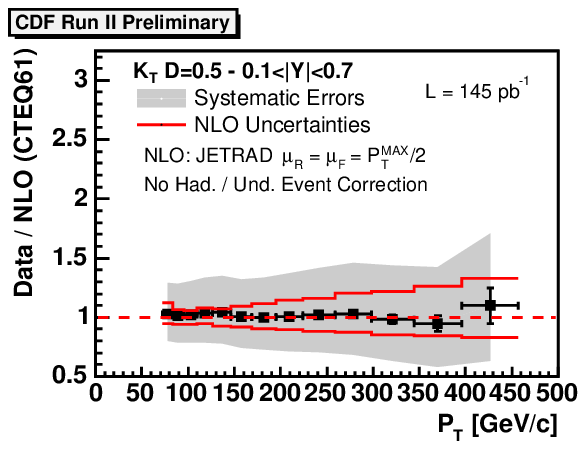 hscale=1 vscale=1  }
\hspace{5.5 cm}
    \special{epsfile=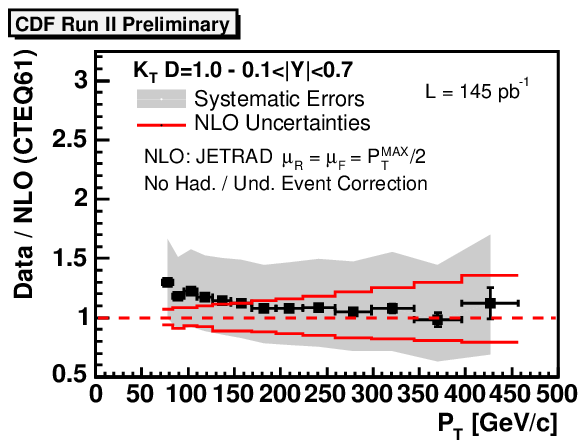 hscale=1 vscale=1  }   
    }}

\vspace{4 cm}
 \caption{\it
    Ratio between the measured inclusive jet cross section and the pQCD NLO predictions 
using the $K_T$ algorithm with D=0.5 and D=0.7, respectively.
    \label{fig2} }
\end{figure}

\begin{figure}[htbp]
  \centerline{\hbox{ \hspace{-6 cm}
    \special{epsfile=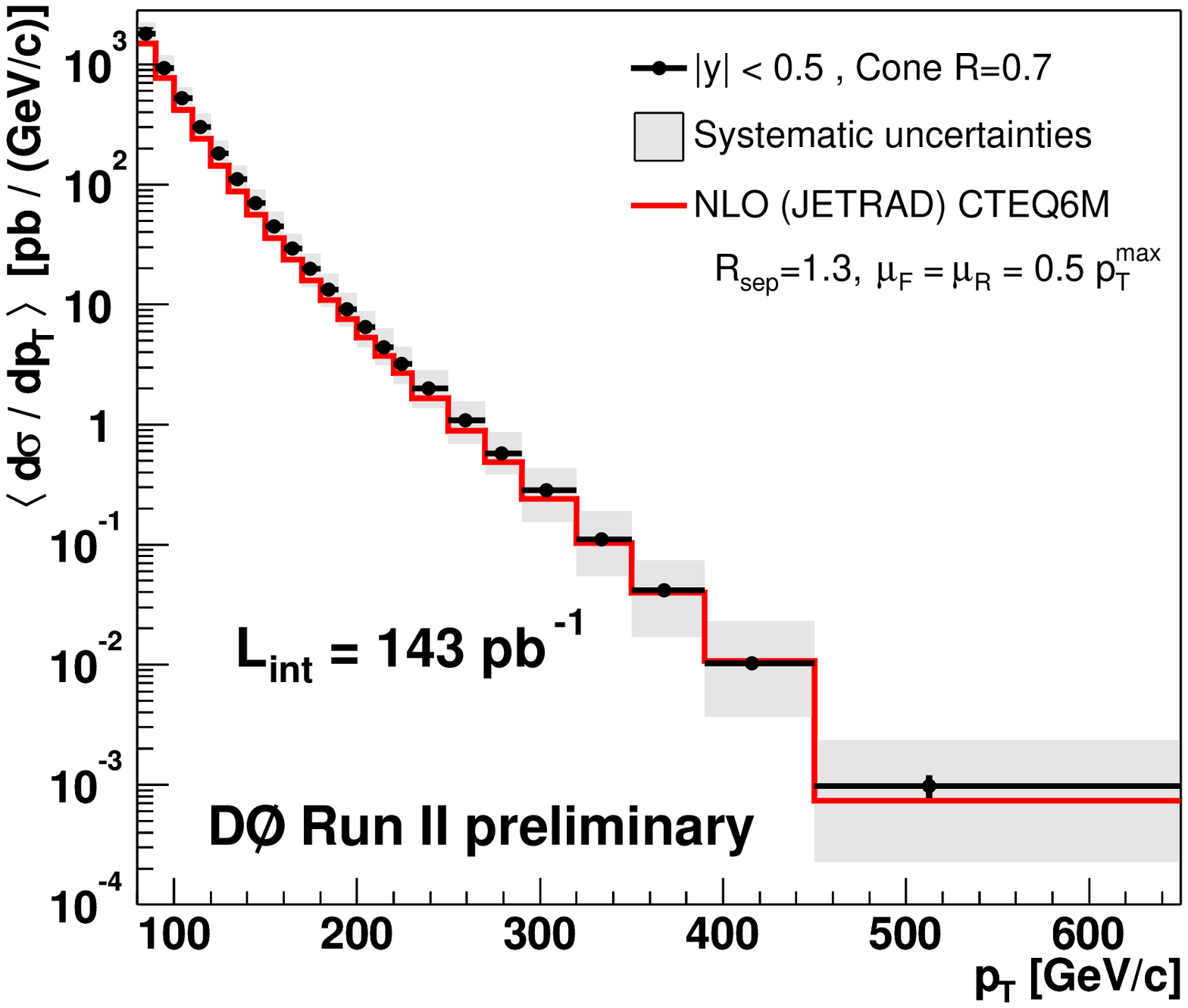 hscale=0.3 vscale=0.3  }   
\hspace{5.5 cm}
    \special{epsfile=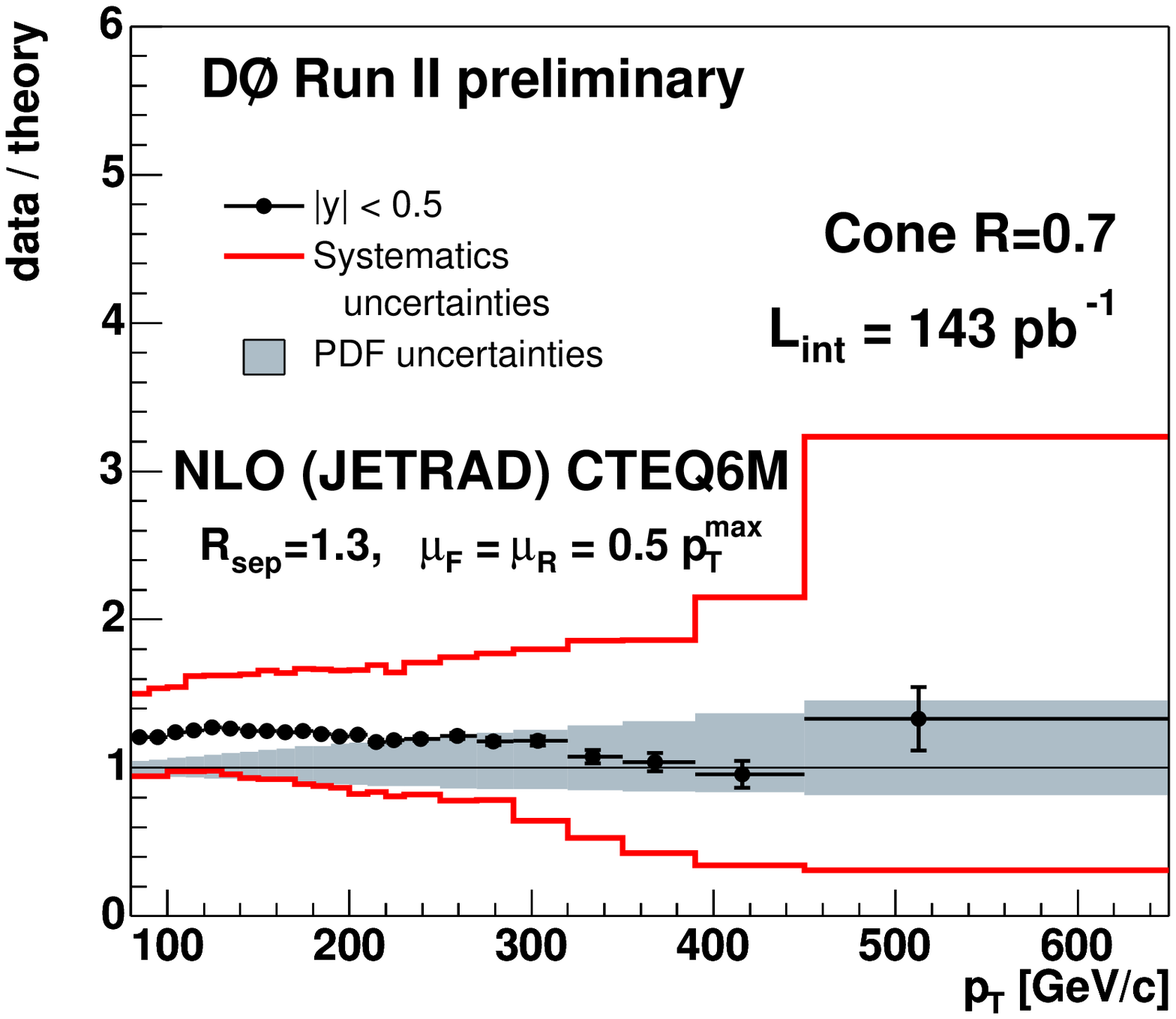 hscale=0.3 vscale=0.3  }   
    }
  }
\vspace{4.7 cm}
 \caption{\it
             The measured inclusive jet cross section by D0 
compared to pQCD NLO predictions. Jets  are seached for using the midpoint jet algorithm. 
    \label{fig3} }
\end{figure}

\begin{figure}[htbp]
  \centerline{\hbox{ \hspace{-6 cm}
    \special{epsfile=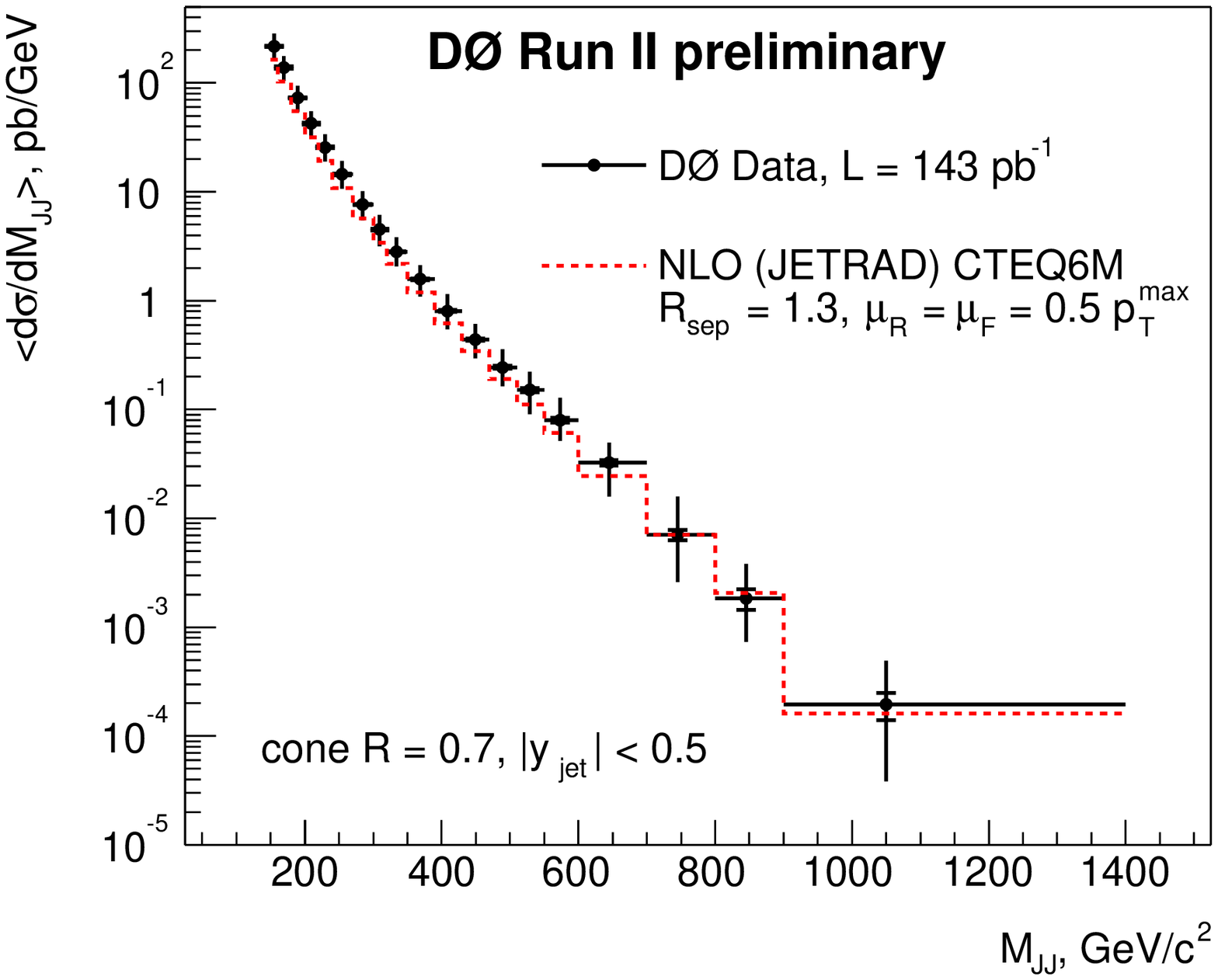   hscale=0.3 vscale=0.3  }
\hspace{5.5 cm}   
    \special{epsfile=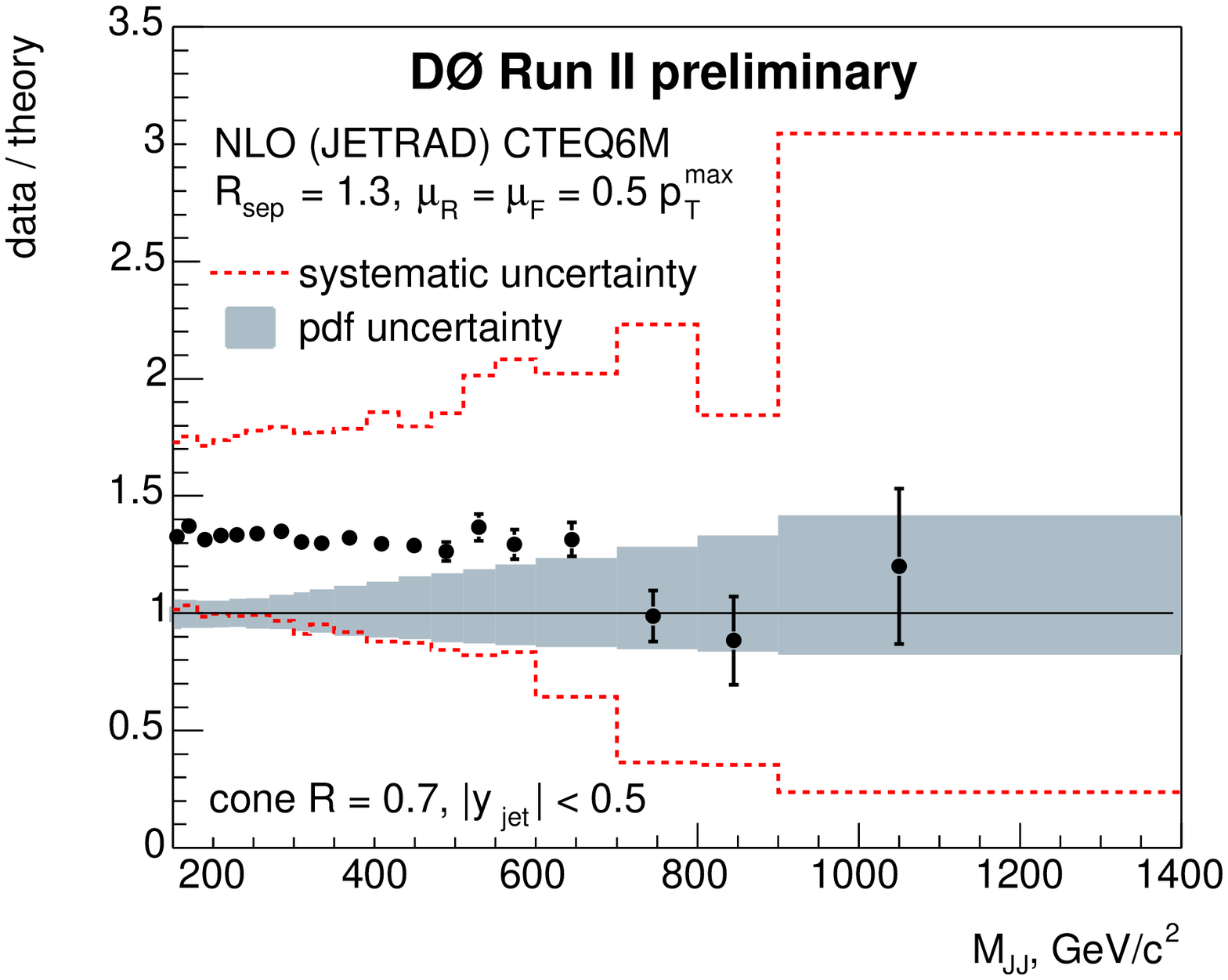 hscale=0.3 vscale=0.3  }   
    }
  }

\vspace{4.5 cm}
 \caption{\it
 The measured inclusive dijet cross section by D0 as a function of the dijet mass  
compared to pQCD NLO predictions.       
    \label{fig4} }
\end{figure}

Figure~\ref{fig3} shows the measured inclusive jet cross section by D0 
based on the first 143$\rm{pb}^{-1}$ of Run II data. The new  
midpoint\cite{midpoint} jet algorithm has been used with a cone size R=0.7. This algorithm 
constitutes an improved version of the cone-based algorithm used in Run I and it is shown to be 
infrared safe when used in fixed-order parton-level calculations.    
The data is in good agreement with the  pQCD NLO predictions using 
CTEQ6 parton density functions and $R_{\rm sep} = 1.3$. However, the measurement 
is dominated by a relatively large uncertainty on the absolute jet energy scale. 
Figure~\ref{fig4} shows the measured cross section by D0 
as a function of the dijet invariant mass in dijet production of central jets. This measurement
is particularly sensitive to the presence of narrow resonances 
decaying into jets of hadrons up to masses of 1.3 TeV. The data is well described by pQCD NLO predictions.  

\begin{figure}[htbp]
  \centerline{\hbox{ \hspace{-6 cm}
    \special{epsfile=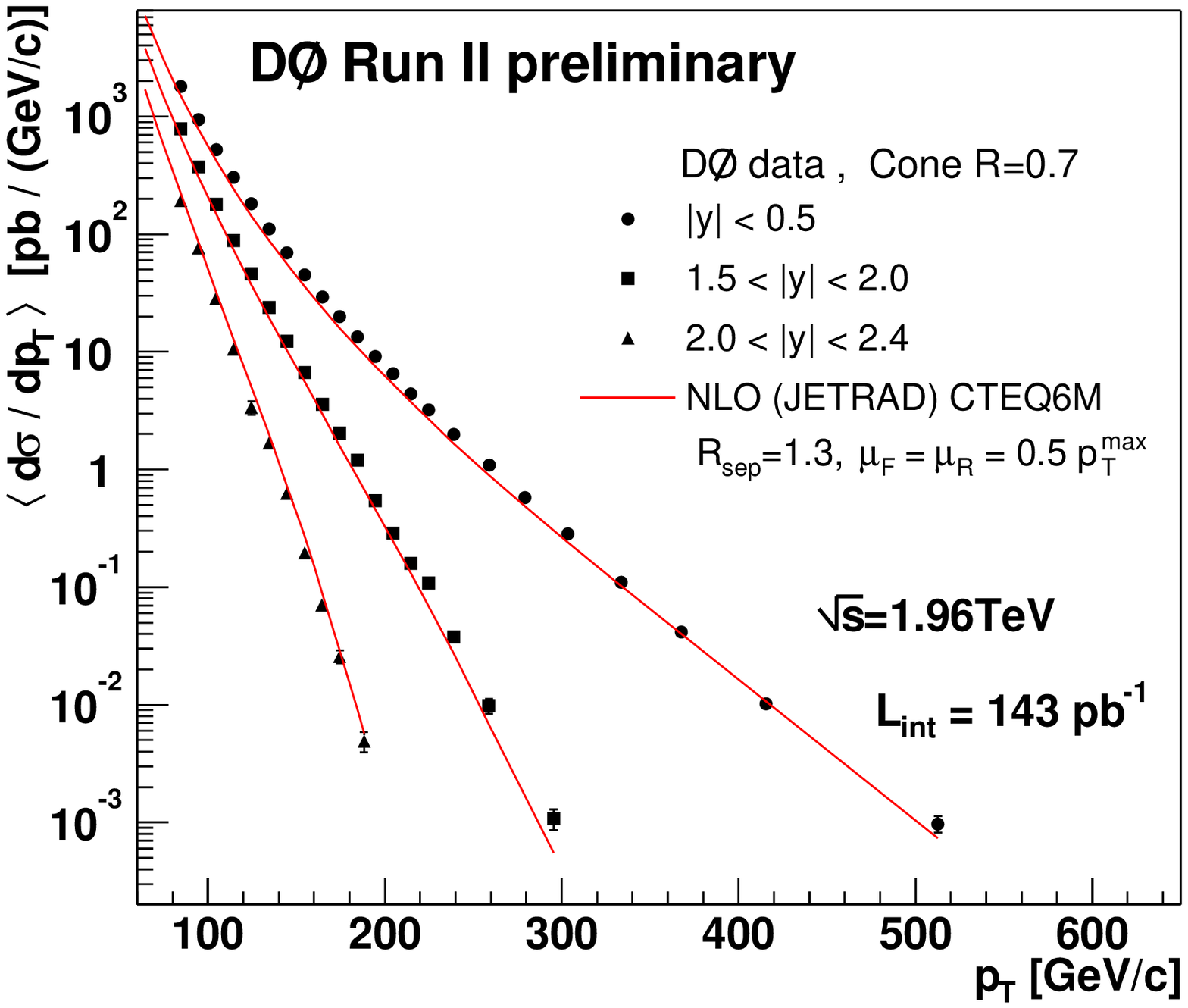 hscale=0.3 vscale=0.3  }   
\hspace{5.5 cm}
    \special{epsfile=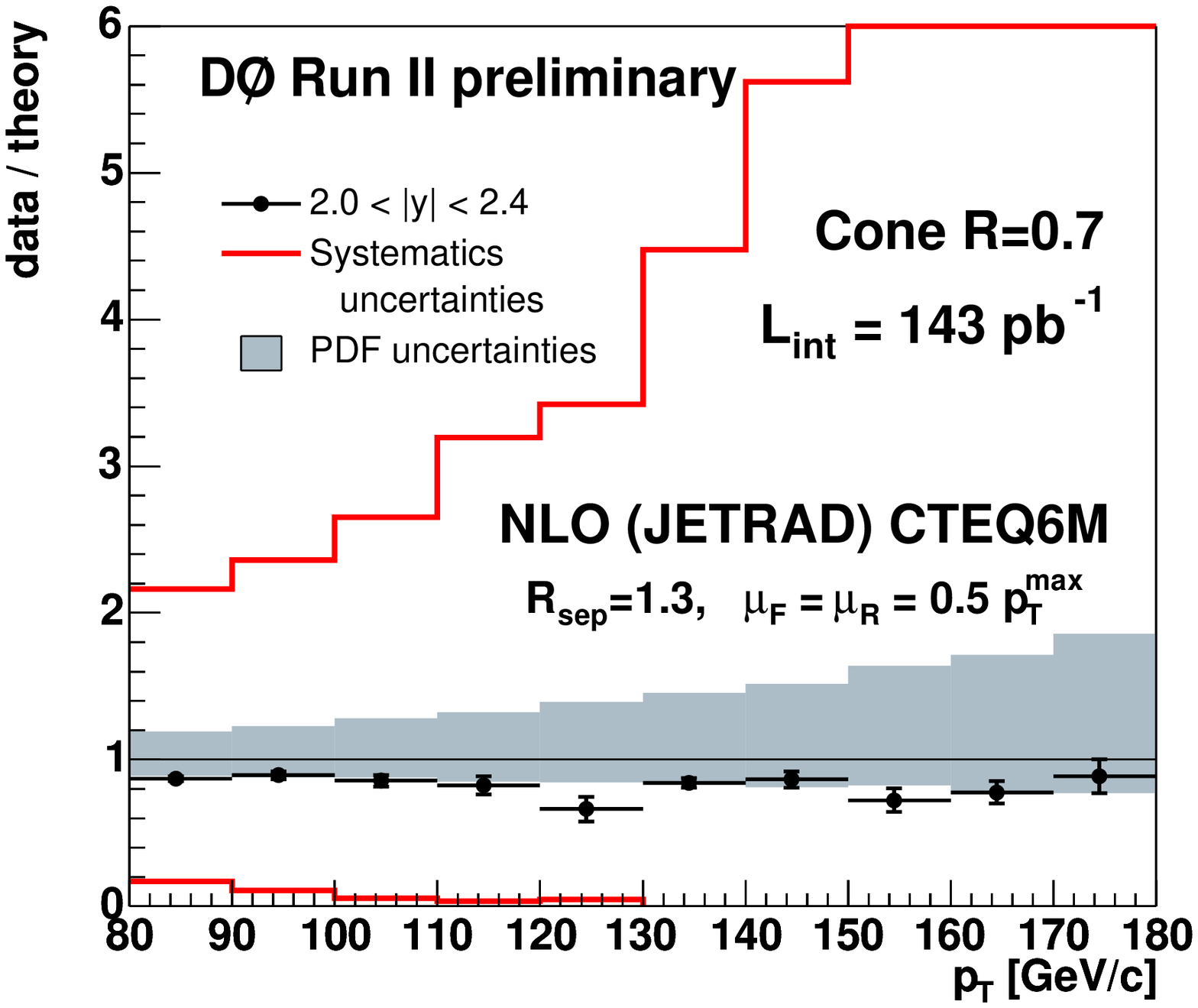  hscale=0.3 vscale=0.3  }   
    }
  }

\vspace{4.5 cm}
 \caption{\it
      (Left) measured inclusive jet cross section by D0 in different regions of rapitidy 
compared to pQCD NLO predictions.(Right)  ratio between the measurements 
and the pQCD NLO predictions for jets with $2.0 <|y| < 2.4$.
    \label{fig5} }
\end{figure}

Nowadays, the Tevatron high-$P^{\rm jet}_T$ jet data
is used, together with prompt-photon data
from fixed target experiments, to constrain the gluon
distribution at high-$x$.
Jet measurements  at large rapidities are important because  
they constrain the gluon density in a region in 
$P_T^{\rm jet}$ where no effect from new physics is expected.       
The D0 experiment has already extended the jet cross section measurements to
the forward region  for jets with $|y| < 2.4$ (see Figure~\ref{fig5}).
At the moment, the results are affected by large systematic errors.
In the near future the experiments will highly reduce their uncertainties and 
precise cross section measurements will allow to further 
constrain the gluon distribution and thus enhance their  
sensitivity to new physics at very high $P_T^{\rm jet}$.
%
%
\section{Study of the Underlying Event}

\begin{figure}[htbp]
  \hbox{ 
  \special{epsfile=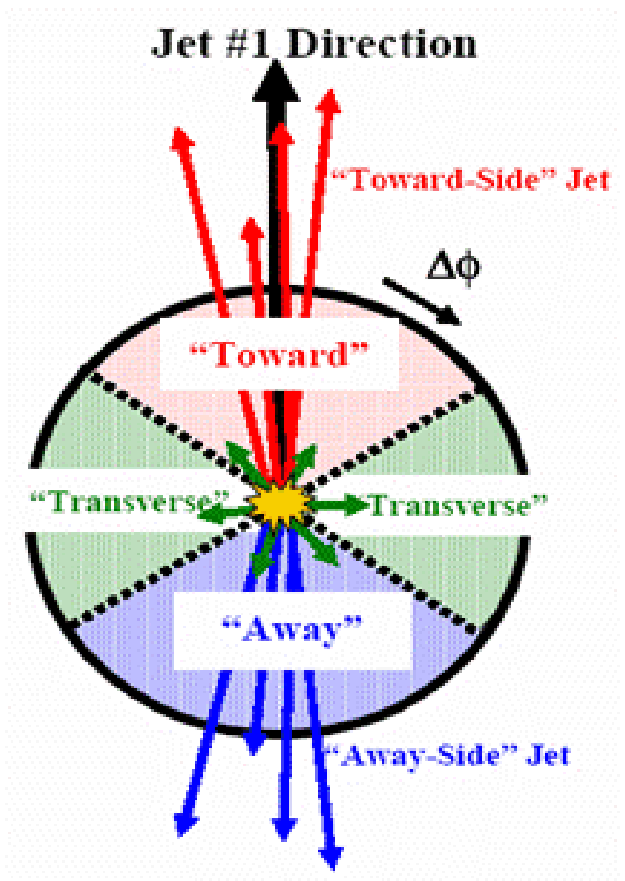 hscale=0.55 vscale=0.55  }
\hspace{5 cm}
  \special{epsfile=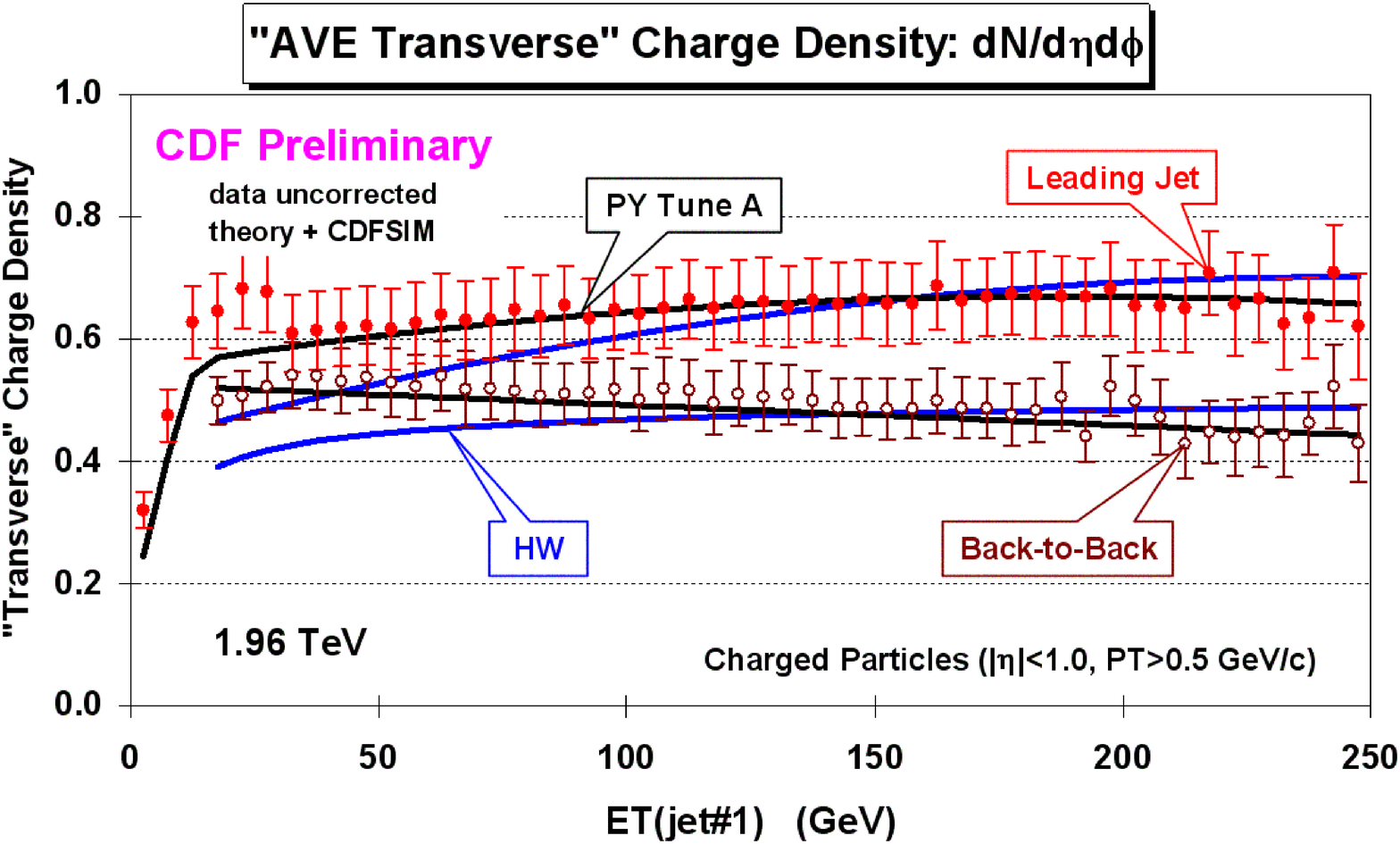 hscale=0.15 vscale=0.2  }
   }

\vspace{5 cm}
\caption{\it (Left) Scheme of the different $\phi$ regions defined
around the leading jet. (Right) Measured average track density
in the transverse region as a function of the $E_T^{\rm jet}$
of the leading jet. The measurements are compared to different Monte Carlo models.   
 \label{fig6} }
\end{figure}

As mentioned in previous section, the hadronic final states at the 
Tevatron are characterized by the presence of soft underlying
emissions, usually denoted as {\it{underlying
 event}}, in addition to highly energetic jets coming from the
hard interaction. The underlying event contains contributions from
initial- and final-state soft gluon radiation, secondary semi-hard partonic
interactions and interactions between the proton and anti-proton remnants that
cannot be described by perturbation theory. These processes  must be
approximately modeled using  Monte Carlo programs tuned to describe the data.
The jet energies measured in the detector contain an underlying event
contribution that has to be subtracted in order to compare the measurements
to pQCD predictions.
Hence, a proper understanding of this underlying event contribution
is crucial to reach the desired precision in
the measured jet cross sections.
In the analysis  presented here, the underlying
event  in dijet
production has been studied by looking at regions well separated from the leading
jets, where the underlying event contribution is expected to dominate the
observed hadronic activity. Jets have been reconstructed using  tracks with
$p_T^{\rm track} > 0.5$ GeV and $|\eta^{\rm track}|<1$ and a cone
 algorithm with R=0.7.
The $\phi$ space around the leading jet is divided
in three regions: {\it{towards}}, {\it{away}}
and {\it{transverse}} (see Figure~\ref{fig6}-left), and the transverse
region is assumed to reflect the underlying event contribution.
Figure~\ref{fig6}-right shows the average track density in the
transverse region as a function of $E_T^{\rm jet}$ of the leading jet for 
the dijet inclusive sample and for events where the leading jets are forced 
to be back-to-back in $\phi$, in order to further reduce extra hard-gluon radiation.
The observed plateau indicates that the underlying event activity
is, to a large extend, independent from the hard interaction.
The measurements have been compared to the
predictions from PYTHIA~\cite{pythia} and HERWIG~\cite{herwig} Monte Carlo programs 
including  leading-order QCD matrix elements plus 
initial and final parton showers. The PYTHIA samples have been created using a special tuned 
set of parameters, denoted as PYTHIA-Tune A, which includes an enhanced contribution from initial-state soft gluon radiation and a tuned set of parameter to control
secondary parton interactions. It was determined as a
result of similar studies of the underlying event performed using 
CDF Run I data \cite{underlying}. PYTHIA-Tune A describes the hadronic activity 
in transverse region  while HERWIG underestimates the radiation at low $E_T^{\rm jet}$.
Similar measurements  in Z+jet(s) events would allow to explore the univesality of the 
underlying event contribution in events with a very different colour configuration in the final state.

\section{Jet Shapes}

The internal structure of jets is dominated by  multi-gluon emissions from the primary final-state parton. 
It is sensitive to the relative quark- and gluon-jet fraction  and 
receives contributions from soft-gluon initial-state radiation and beam remnant-remnant interactions.
The study of jet shapes at the Tevatron provides a stringent test of QCD predictions  and tests 
the validity of the models for parton cascades and soft-gluon emissions in hadron-hadron collisions. 
The CDF experiment has presented results on jet shapes for 
central jets with transverse momentum in the region $37$ GeV $< P_T^{\rm jet} < 380$~GeV, 
where jets are searched for using the midpoint\footnote{A $75 \%$ 
merging fraction has been used instead of the default $50 \%$.} algorithm  and a cone size $R=0.7$.
The integrated jet shape, $\Psi(r)$, is defined as the average fraction of the 
jet transverse momentum that lies inside a cone of radius $r$ concentric to the jet cone:
\begin{equation}
\Psi(r) = \frac{1}{\rm N_{jet}} \sum_{\rm jets} \frac{P_T(0,r) }{P_T(0,R)}, \ \ \ \ 0 \leq  r \leq R
\end{equation}
\noindent
where $N_{\rm jet}$ denotes the number of jets. The measured jet shapes have been compared to the
predictions from PYTHIA-Tune A and HERWIG Monte Carlo programs.

\begin{figure}[h]
\vspace{-0.5 cm}
  \centerline{\hbox{ \hspace{-6 cm}
    \special{epsfile=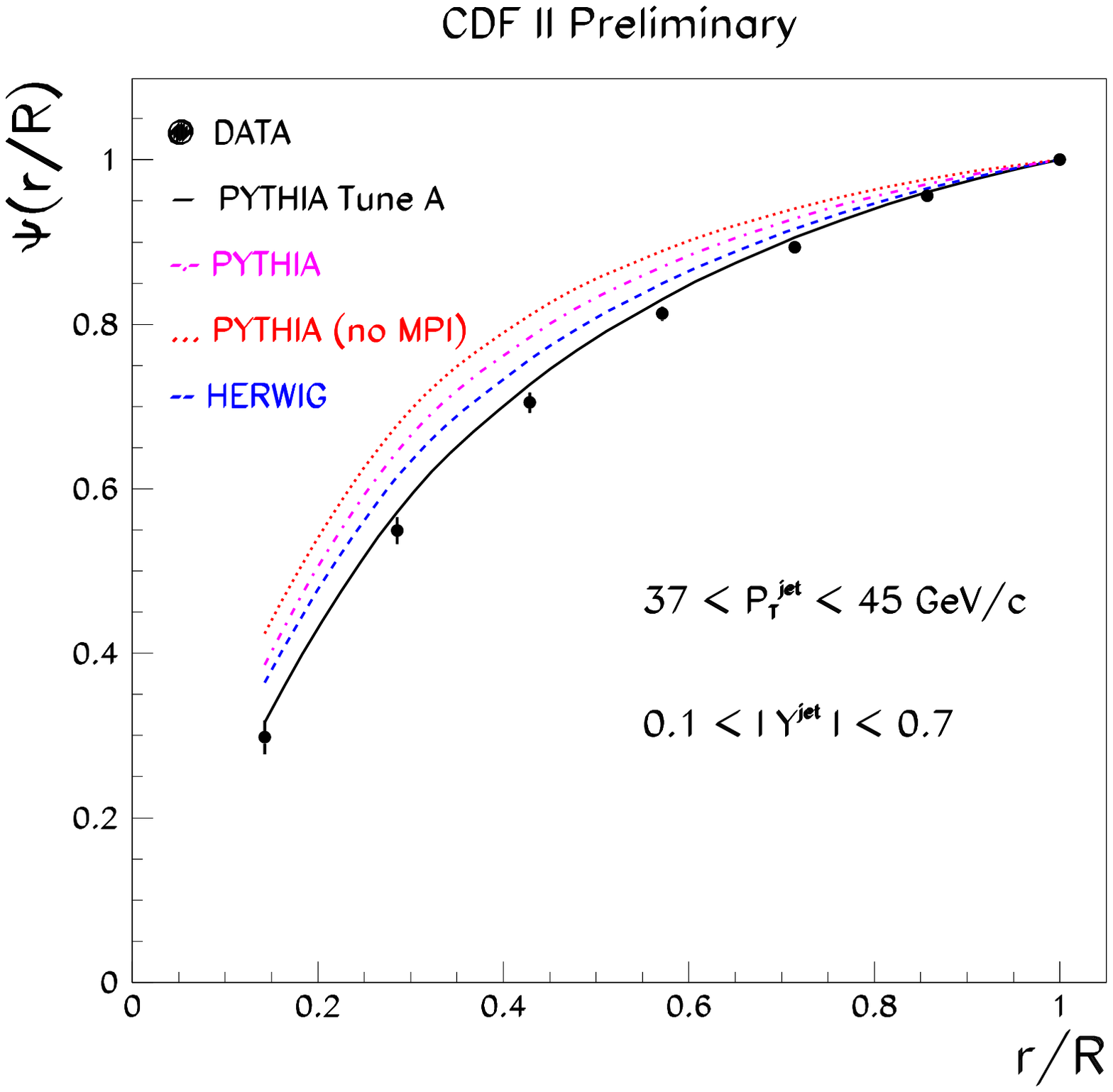 hscale=0.285 vscale=0.285  }    
    \hspace{5  cm} 
    \special{epsfile=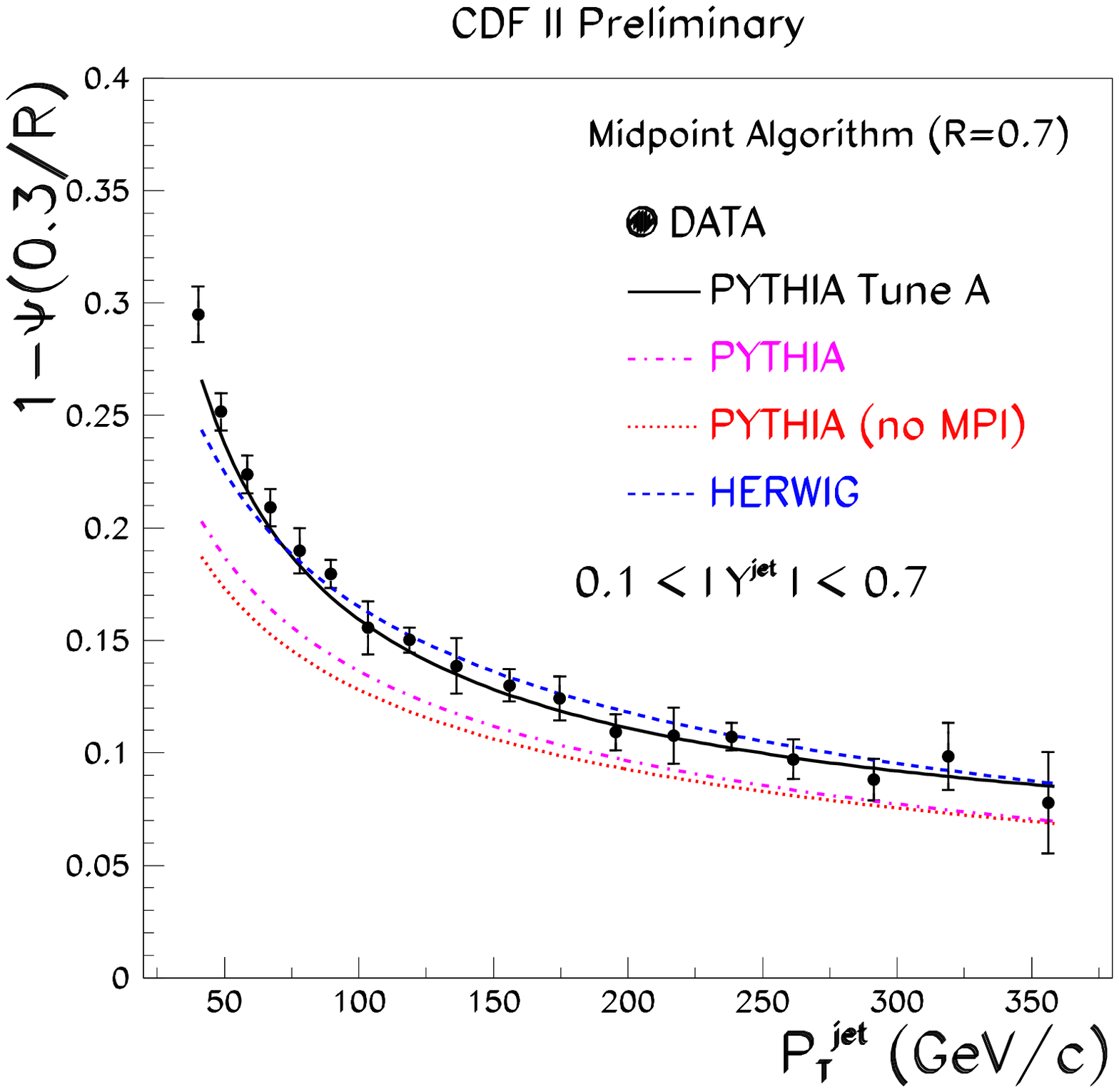            hscale=0.285 vscale=0.285  }
    }
  }
\vspace{4.9 cm}
\caption{\it The measured integrated jet shape compared to different Monte Carlo predictions.
    \label{fig7} }
\end{figure}

\noindent
In addition, two different PYTHIA samples have been used with default parameters and with and without
the contribution from multiple parton interactions (MPI) between proton and antiproton remnants, the latter
denoted as PYTHIA-(no MPI), to illustrate the importance of a proper modeling of soft-gluon
radiation in describing the measured jet shapes. Figure~\ref{fig7}(left)  presents the measured integrated jet shapes, $\Psi(r/R)$, 
for jets with $37 < P_{T}^{\rm jet} < 45$ GeV, compared to
HERWIG, PYTHIA-Tune A,  PYTHIA and PYTHIA-(no MPI) predictions. In addition, Figure~\ref{fig7}(right) 
shows, for a fixed radius $r_0 = 0.3$, the average
fraction of the jet transverse momentum outside $r=r_0$, $1-\Psi(r_0/R)$, as a function of
$P_{T}^{\rm jet}$ where the points are located at the weighted mean in each $P_{T}^{\rm jet}$ range.
The measurements show that the fraction of jet transverse momentum at a given fixed $r_0/R$ increases 
($1-\Psi(r_0/R)$ decreases) with $P_{T}^{\rm jet}$, indicating 
that the jets become narrower as  $P_{T}^{\rm jet}$ increases. PYTHIA with default parameters
produces jets systematically narrower than the data in the whole region in $P_{T}^{\rm jet}$. The contribution from
secondary parton interactions between remnants to the predicted jet shapes 
(shown by the difference between  PYTHIA and PYTHIA-(no MPI) predictions) is relatively small 
and decreases as $P_{T}^{\rm jet}$ increases. PYTHIA-Tune A predictions describe all of 
the data well. HERWIG predictions describe
the measured jet shapes well for  $P_{T}^{\rm jet} > 55$ GeV  but
produces jets that are too narrow at lower $P_{T}^{\rm jet}$.  
\begin{figure}[h]
\vspace{-0.5cm}
  \centerline{\hbox{
    \hspace{-6 cm} 
    \special{epsfile=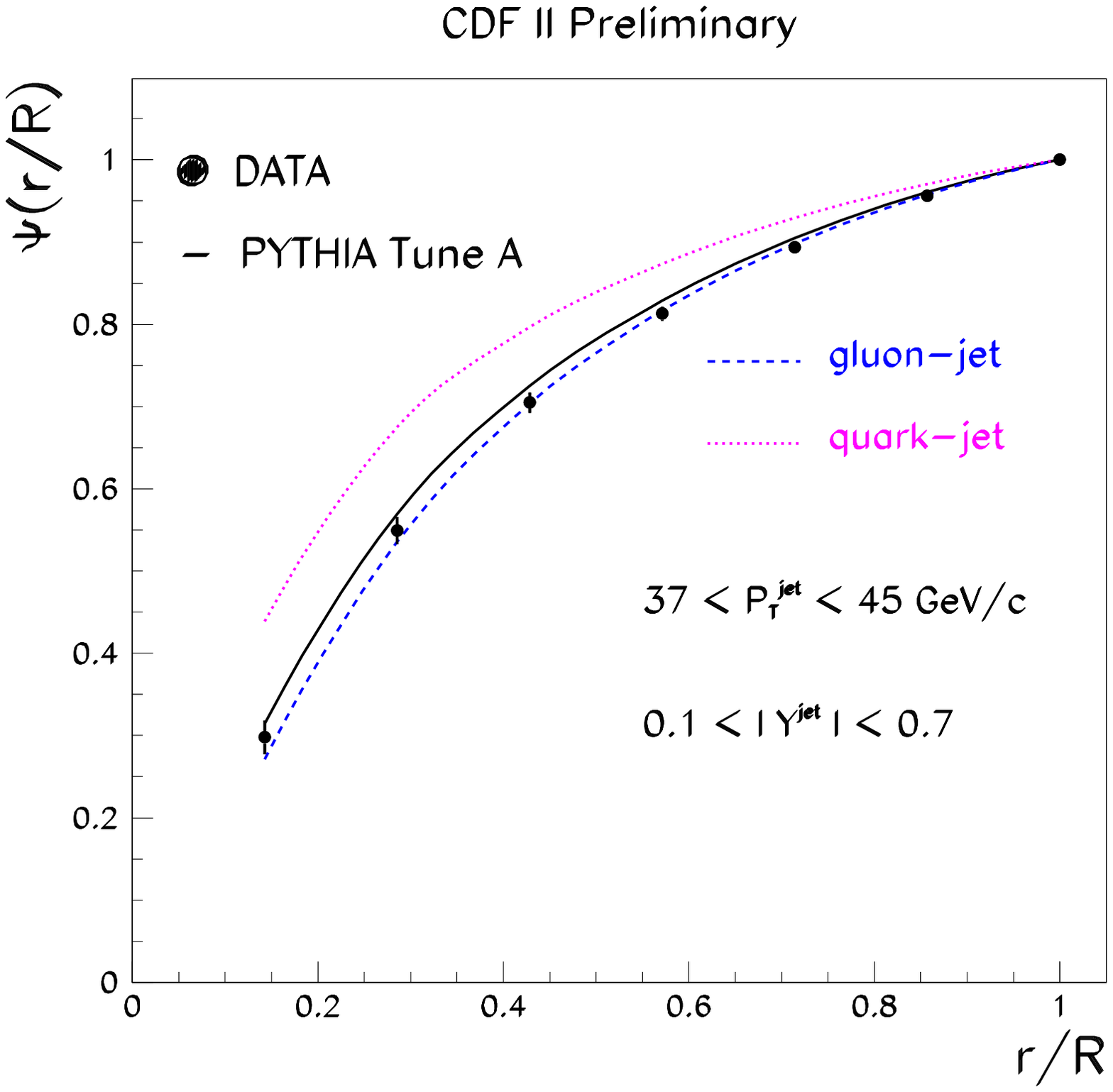 hscale=0.285 vscale=0.285  }    
    \hspace{5  cm} 
    \special{epsfile=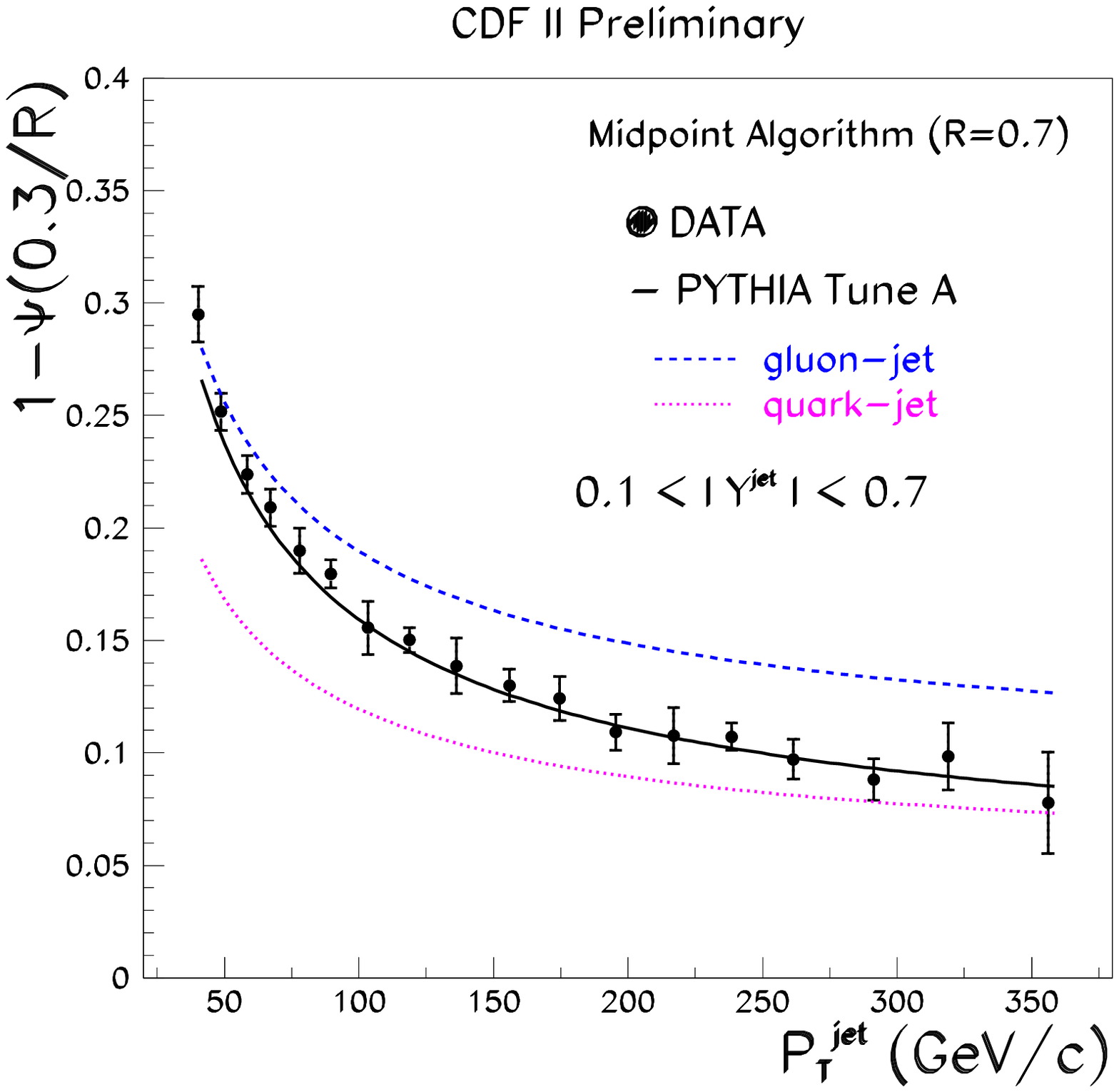            hscale=0.285 vscale=0.285  }
    }
  }

\vspace{4.9 cm}
\caption{\it The measured integrated jet shape compared to the predictions of 
PYTHIA-Tune A and the separated contributions from quark-  
and gluon-jets.
    \label{fig8} }
\end{figure}
Figure~\ref{fig8}(left) shows the measured integrated jet shapes, $\Psi(r/R)$, for jets with 
$37 < P_{T}^{\rm jet} < 45$ GeV, compared to PYTHIA-Tune A  and  the predictions for
quark- and gluon-jets\footnote{Each hadron-level jet from PYTHIA is classified as a quark- or gluon-jet 
by matching ($y-\phi$ plane) its
directions with that of one of the outgoing partons from
the hard interaction.} separately. Figure~\ref{fig8}(right) shows the measured $1-\Psi(r_0/R)$, $r_0 = 0.3$, as a function of
$P_{T}^{\rm jet}$. The Monte Carlo predictions indicate that the measured jet shapes are dominated by
contributions from gluon-initiated jets at low $P_{T}^{\rm jet}$ while contributions from
quark-initiated jets become important at high $P_{T}^{\rm jet}$. This can be explained in terms of
the different partonic contents in the proton and antiproton  in the low- and high-$P_{T}^{\rm jet}$ regions, since the
mixture of gluon- and quark-jet in the final state  partially reflects
the nature of the incoming partons that participate in the hard interaction.
For a given type of parton-jet in the Monte Carlo (quark- or gluon-jet), the observed trend with  $P_{T}^{\rm jet}$
shows the running of the strong coupling constant, $\alpha_s(P_{T}^{\rm jet})$. Jet shape measurements 
thus introduce strong constrains on phenomenological models describing soft-gluon 
radiation and the underlying event in hadron-hadron interactions. Similar studies with b-tagged jets
will be necessary to test our knowledge of b-quark jet fragmentation processes in
hadronic interactions, which is essential for future precise Top and Higgs
measurements.
%
\section{$\Delta \phi_{\rm dijet}$ Decorrelations}

The D0 experiment has employed the dijet sample to study azimuthal decorrelations, $\Delta \phi_{\rm dijet}$,  between the 
two leading jets. The normalized cross section:
\begin{equation}
\frac{1}{\sigma_{\rm dijet}} \frac{d \sigma}{d \Delta \phi_{\rm dijet}}
\end{equation}
is sensitive to the spectrum  of the gluon radiation in the event. The measurements has been performed in 
different regions of the leading jet $P_T^{\rm jet}$ starting at $P_T^{\rm jet} > 75$~GeV and the second jet is 
requied to have at least $P_T^{\rm jet} > 40$~GeV.

\begin{figure}[htbp]
  \centerline{\hbox{ \hspace{-6 cm}
    \special{epsfile=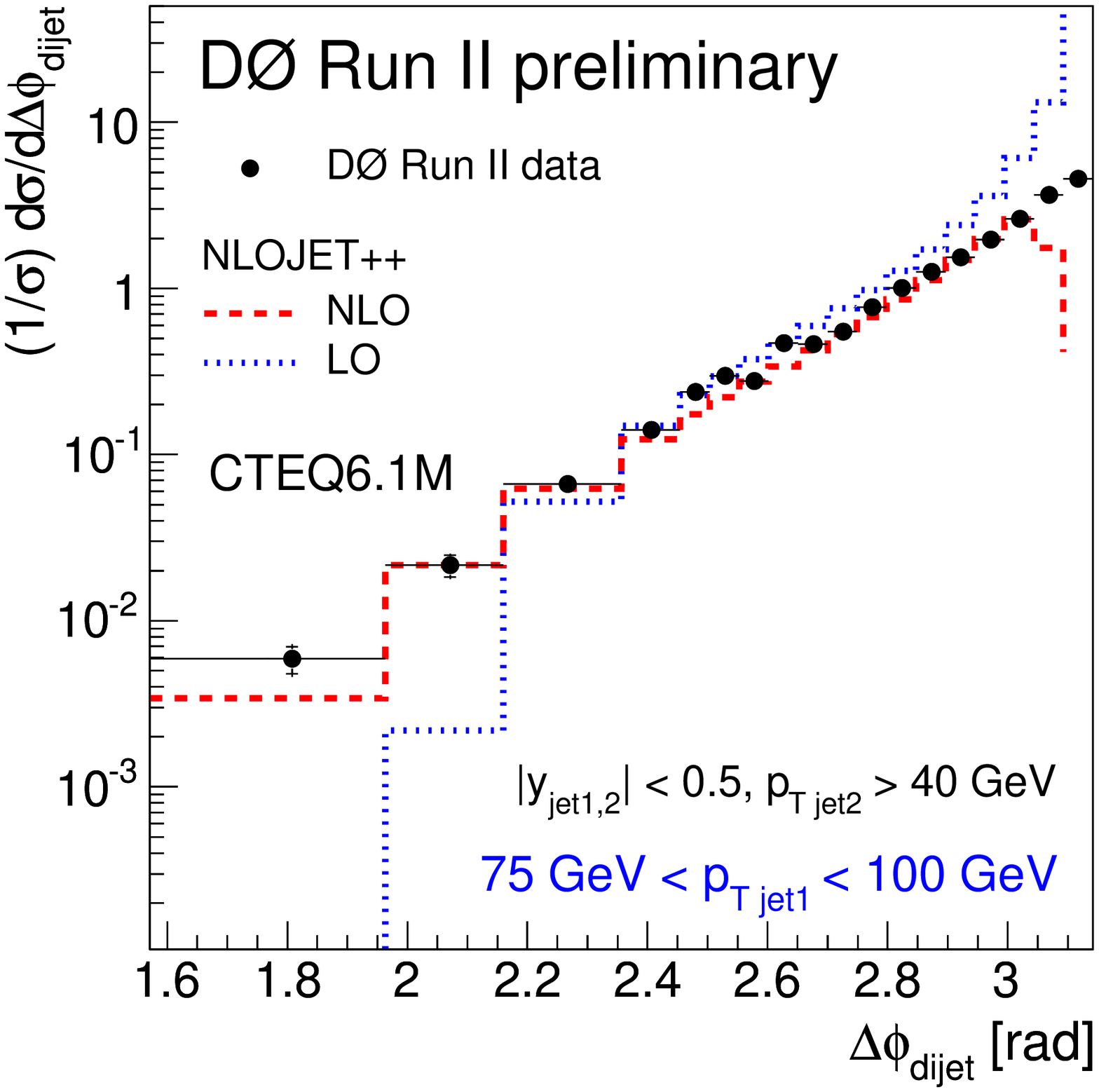 hscale=0.275 vscale=0.28  }
\hspace{5.5 cm}  
    \special{epsfile=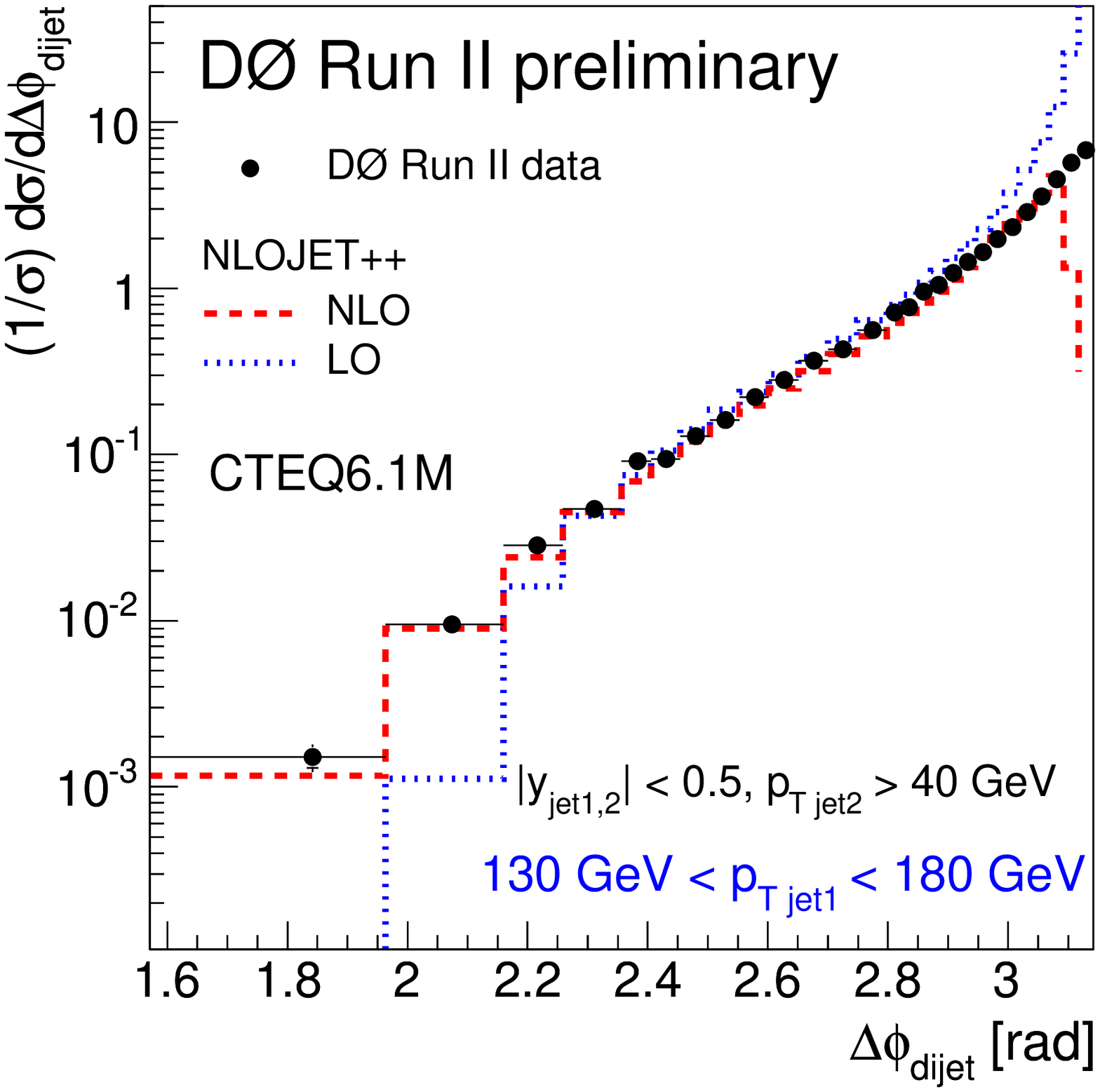 hscale=0.275 vscale=0.28  }  
    }
  }

\vspace{5.2 cm}
 \caption{\it
      Measured azimuthal decorrelations in dijet production for central jets compared to pQCD predictions in 
different regions of $P^{\rm jet}_T$ of the leading jet.
    \label{fig9} }
\end{figure}%

Figure~\ref{fig9} shows the measured cross section compared to LO and NLO predictions from NLOJET++ program~\cite{nlojet++}. 
The LO predictions, with at most three partons in the final state, is limited to $\Delta \phi_{\rm dijet} > 2 \pi/3$, 
for which the three partons define a {\it{Mercedes-star}} topology. It presents a prominent peak 
at $\Delta \phi_{\rm dijet} = \phi$ corresponding to the soft limit for which the third parton is collinear 
to the direction of the two leading partons. The NLO predictions, with four partons in the final state, describes 
the measured  $\Delta \phi_{\rm dijet}$ distribution except at very high and very low values of  $\Delta \phi_{\rm dijet}$
where additional soft constributions, corresponding to a resummed calculation, are necessary. A reasonable approximation 
to such calculations is provided by parton shower Monte Carlo programs.

\begin{figure}[htbp]
  \centerline{\hbox{ \hspace{- 6 cm}
    \special{epsfile=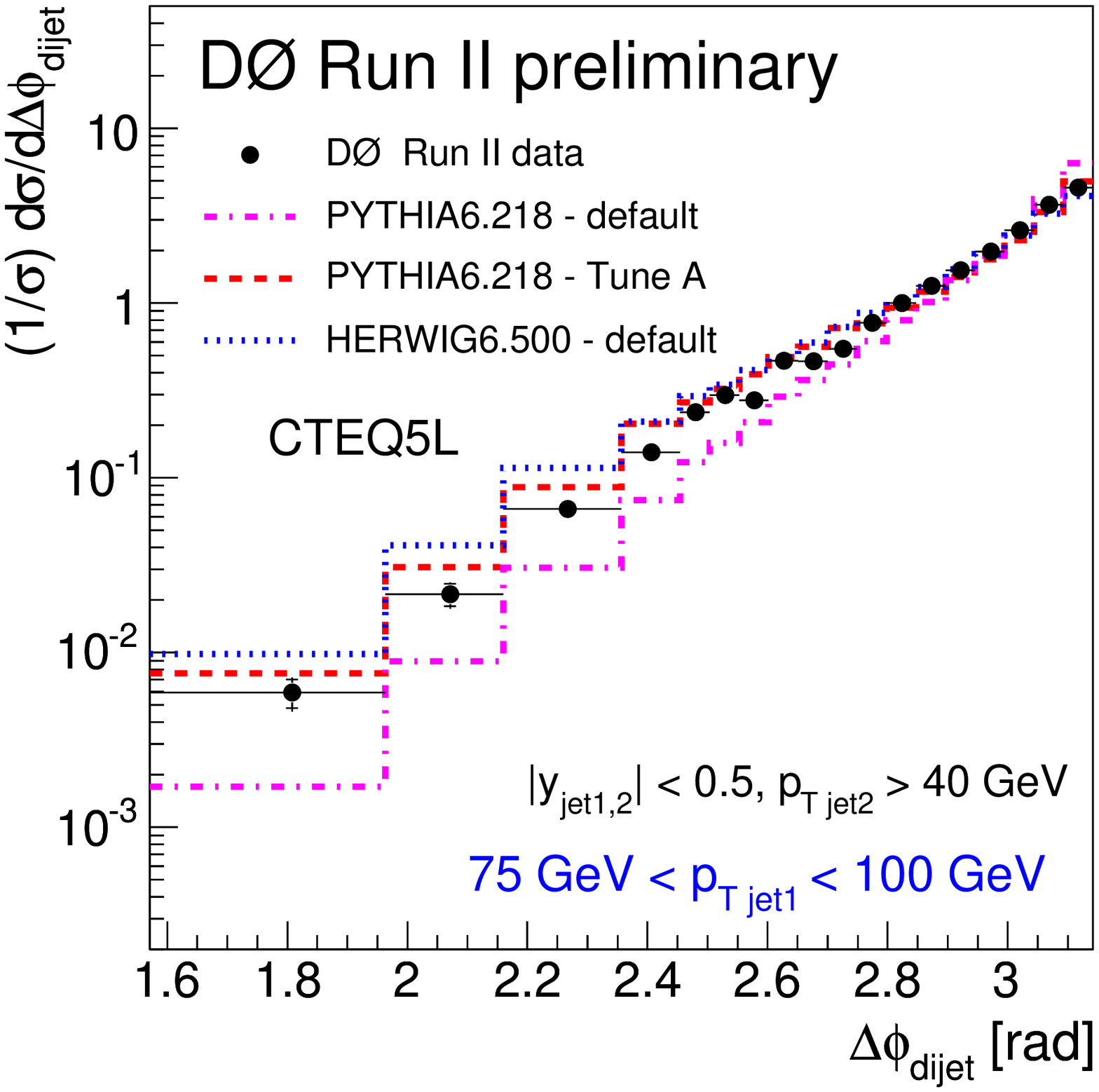 hscale=0.285 vscale=0.275  }  
\hspace{5.5 cm}
    \special{epsfile=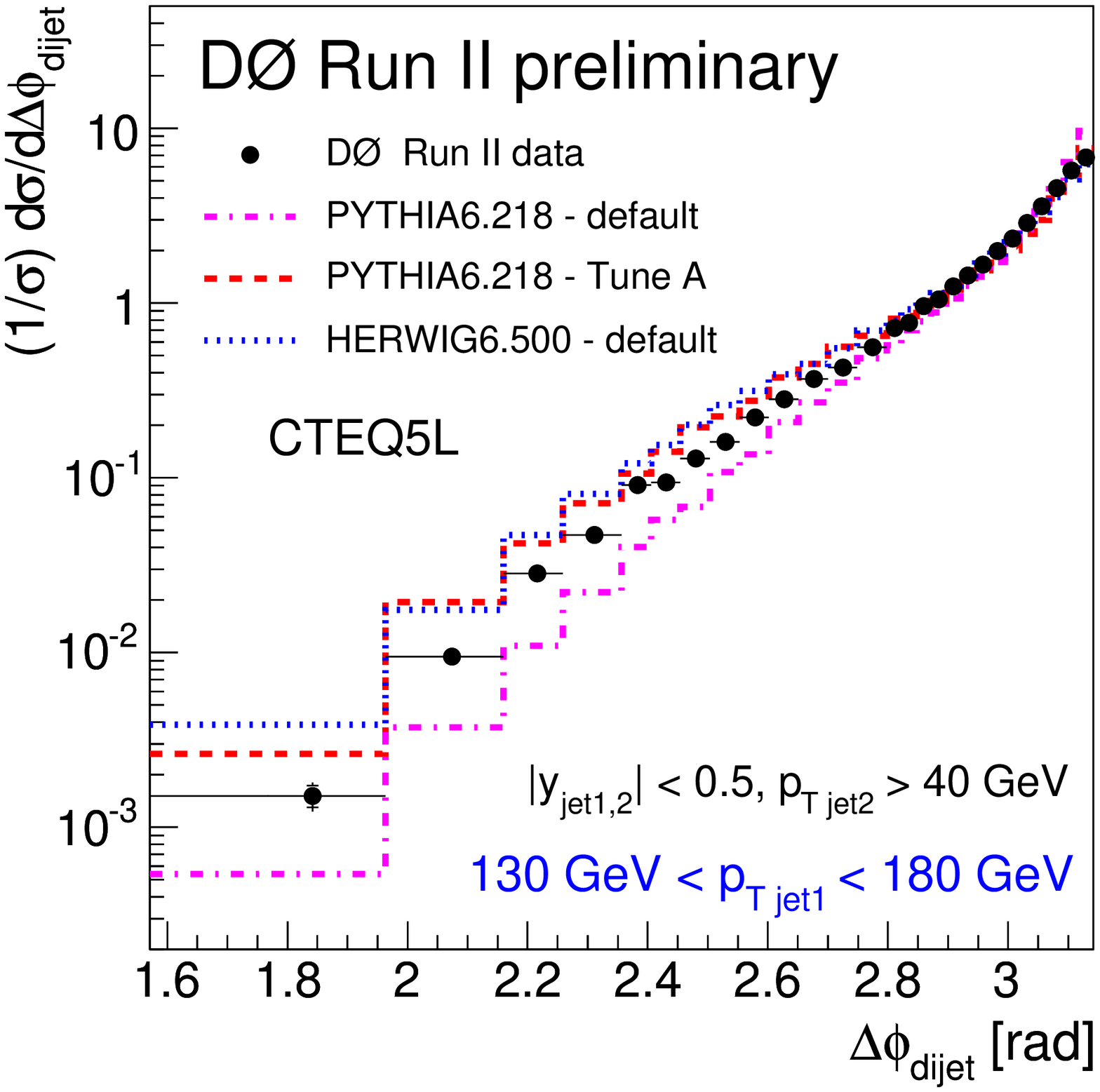 hscale=0.285 vscale=0.275  }  
    }
  }

\vspace{5 cm}
 \caption{\it
   Measured azimuthal decorrelations in dijet production for central jets compared to PYTHIA and HERWIG
 predictions in different regions of leading $P^{\rm jet}_T$.     
    \label{fig10} }
\end{figure}

Figure~\ref{fig10} presents the measured cross section compared to PYTHIA-Tune A, PYTHIA and HERWIG predictions 
in different regions of $P_T^{\rm jet}$. PYTHIA with default parameters underestimates the gluon radiation at large angles.
PYTHIA-Tune A predictions, which include an enhanced contribution from initial-state soft gluon radiation and secondary parton interactions, describe the azimuthal distribution. HERWIG also describes the data although tends to produce 
less radiation than PYTHIA-Tune A close to the direction of the leading jets.

\section{Summary and Conclusions}
Both CDF and $\D$ experiments have carried out measurements of the inclusive jet production 
cross section in Run II at the Tevatron using different jet algorithms. The measurements
are in agreement with pQCD NLO predictions. Dedicated studies of the underlying event, jet shapes
and azimuthal decorrelations in dijet final states  allowed to establish the validity of the 
Monte Carlo models used to describe the soft-gluon contributions in the final state. 
\section{Acknowledgements}

I thank the organizers for their kind invitation to the conference and 
for the exciting program of talks and discussions they made possible.

%

%


\end{document}